\documentclass[conference,a4paper]{IEEEtran}
\IEEEoverridecommandlockouts
\usepackage{cite}
\usepackage{amsmath,amssymb,amsfonts}
\usepackage{graphicx}
\usepackage{xcolor}
\usepackage{booktabs}
\usepackage{array}
\usepackage{multirow}
\usepackage{url}
\usepackage{microtype}
\usepackage{placeins}
\usepackage{tikz}
\usetikzlibrary{arrows.meta,positioning,calc}
\usepackage{pgfplots}
\pgfplotsset{compat=1.18}

\newcommand{\KM}{K_{\mathrm{M}}}
\newcommand{\KQ}{K_{\mathrm{Q}}}
\newcommand{\KC}{K_{\mathrm{C}}}
\newcommand{\Kmix}{K_{\alpha}}
\newcommand{\R}{\mathbb{R}}

\definecolor{oiBlue}{HTML}{0072B2}
\definecolor{oiVermillion}{HTML}{D55E00}
\definecolor{oiGreen}{HTML}{009E73}

\begin{document}

\title{QBioFusion-QSAR: Morgan-Anchored Quantum Multiple Kernel Learning for Small-Data Ligand Classification}
\author{\IEEEauthorblockN{Azadeh Alavi\IEEEauthorrefmark{1},
Fatemeh Kouchmeshki\IEEEauthorrefmark{3},
Muhammad Usman\IEEEauthorrefmark{4},
Jessica Holien\IEEEauthorrefmark{2}
}
\IEEEauthorblockA{\IEEEauthorrefmark{1}School of Computing Technologies, RMIT University, Melbourne 3000, Australia}
\IEEEauthorblockA{\IEEEauthorrefmark{2}School of Biology, STEM College, RMIT University, Melbourne 3000, Australia}
\IEEEauthorblockA{\IEEEauthorrefmark{3}Pattern Recognition Pty Ltd, Melbourne 3240, Australia}
\IEEEauthorblockA{\IEEEauthorrefmark{4}Data61, CSIRO, Clayton, VIC, Australia}
\IEEEauthorblockA{*Corresponding author: Azadeh Alavi and Jessica Holein}}

\maketitle

\begin{abstract}
Small quantitative structure-activity relationship (QSAR) studies are difficult when close molecular analogues have different activity labels. This paper asks whether a quantum kernel can add similarity information to a Morgan/Tanimoto fingerprint model, and which molecules account for the change. QBioFusion-QSAR uses quantum multiple kernel learning (QMKL): a support vector machine combines a Morgan/Tanimoto kernel with a quantum fidelity kernel constructed from fold-local components derived from RDKit and Mordred descriptors and Deep-PK features. Linear and radial basis function descriptor kernels are included as classical controls. On the 54-molecule PsychLight-A benchmark, Morgan/Tanimoto was the strongest single representation. In the primary stratified five-fold evaluation, QMKL increased accuracy from 0.815 to 0.833 and Matthews correlation coefficient (MCC) from 0.613 to 0.645. Matched-regularization auditing attributed the change to N-Me-5-HT and N-Me-tryptamine changing from false-negative to true-positive predictions; activity-cliff subset MCC increased from 0.07 to 0.22. Repeating the five-fold protocol over ten random partitionings showed that learned QMKL did not exceed Morgan/Tanimoto on mean MCC; paired held-out bootstrap intervals for the matched comparison also span zero. These results support QBioFusion-QSAR as an auditable QMKL framework for identifying localized residual quantum-kernel contributions in small-data, activity-cliff-aware ligand classification. 
\end{abstract}


\section{Introduction}

Quantitative structure-activity relationship (QSAR) models are used to relate molecular structure to biological activity and to prioritize compounds before experimental characterization. This is especially relevant for serotonin-receptor drug discovery, where ligand structure, receptor selectivity, biased agonism, and pharmacological context interact in ways that are difficult to summarize using a single handcrafted descriptor. Psychedelic and psychoplastogenic ligands are frequently aminergic molecules and are commonly organized around tryptamine, phenethylamine, and ergoline-like structural families. Their pharmacology is closely connected to the 5-hydroxytryptamine 2A (5-HT$_{2A}$) receptor, a class-A G-protein-coupled receptor whose downstream signaling includes G-protein and beta-arrestin pathways \cite{Nichols2016,Kwan2022,Wootten2018}. Recent work has therefore examined machine learning for predicting hallucinogenic potential and serotonergic activity from molecular structure \cite{Urbina2024}.

The present study considers the small-data regime represented by PsychLight-A, a curated ligand-classification dataset with 54 molecules, 18 positives, and 36 negatives. The descriptor set used in this study contains Morgan fingerprints, RDKit descriptors, Mordred descriptors, and Deep-PK predicted pharmacokinetic and toxicity features. This setting is challenging for two related reasons. First, each molecule can be described by many more numerical features than there are labeled examples, which makes model selection sensitive to the particular training split. Second, chemically similar ligands can have different labels. In a QSAR task, this second difficulty is often more important than feature count alone: a fingerprint kernel may place two ligands close to one another because they share substructures, even though small substitutions, stereochemistry, or receptor-context effects produce different activity labels. The evaluation must therefore report held-out performance and also identify the molecules whose predicted class changes.

Quantum kernel methods are a natural quantum machine learning (QML) tool for this setting because they separate representation from prediction. A molecular descriptor vector can be encoded as a quantum state, pairwise molecular similarity can be computed from quantum-state overlap, and a classical kernel classifier can then use the resulting similarity matrix \cite{SchuldKilloran2019,Havlicek2019,Schuld2021}. Multiple kernel learning (MKL) provides the corresponding classical framework for combining different molecular similarity measures in one margin-based classifier. Recent quantum multiple kernel learning (QMKL) work in QSAR similarly frames the classifier as a support vector machine (SVM) trained on an optimized combination of quantum and classical kernels \cite{Giraldo2025}. This is the perspective adopted here: Morgan/Tanimoto similarity supplies the fingerprint-based anchor, and a descriptor-derived quantum fidelity kernel is evaluated as an additional similarity source whose weight is selected only within inner training folds.

The paper is organized around two questions. The first question is whether a descriptor-derived quantum kernel, embedded in a Morgan-anchored MKL formulation, adds a useful similarity pattern beyond the Morgan/Tanimoto anchor. The second question is how the MKL combination changes the classifier: which molecules move across the decision boundary, in which Morgan-neighborhoods those changes occur, and how much of the effect remains after matched classical descriptor-kernel controls. The study therefore separates evidence of a predictive change from molecule-level contribution analysis. The aggregate analysis asks whether the embedded quantum kernel improves the primary held-out evaluation, whereas the contribution analysis asks where the improvement appears and whether it is specific to the quantum fidelity geometry or shared with classical descriptor kernels.

This work makes four contributions. First, it introduces QBioFusion-QSAR, a compact Morgan-anchored QMKL pipeline for small-data ligand classification. Second, it gives a fold-local validation protocol in which imputation, scaling, principal-component analysis, descriptor rescaling, kernel construction, SVM fitting, and score transformation are repeated inside the corresponding training partition. Third, it adds representation ablations for RDKit, Mordred, and Deep-PK fold-local descriptor components, together with matched classical descriptor-kernel controls, so that the quantum fidelity geometry is interpreted against the same reduced descriptor inputs. Fourth, it combines primary five-fold evaluation, matched-regularization comparison, fixed-weight profiling, activity-cliff threshold sensitivity, curation sensitivity, repeated random states, paired uncertainty estimates, and molecule-level auditing to quantify where and to what extent the embedded quantum kernel contributes. The intended claim is therefore focused: Morgan-anchored QMKL exposes a promising, localized residual-similarity effect whose generalization requires larger curated validation.

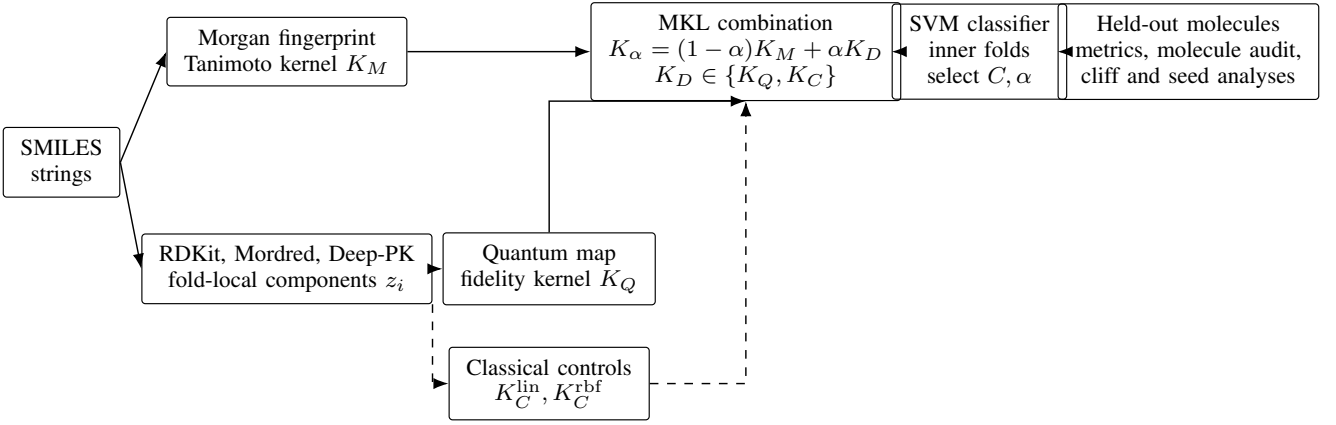
\begin{figure*}[t]
    \centering
    \resizebox{0.96\textwidth}{!}{%
    \begin{tikzpicture}[
        font=\footnotesize,
        block/.style={draw, rounded corners=1.2pt, align=center, minimum height=0.82cm, inner xsep=6pt, inner ysep=4pt},
        arrow/.style={-Latex, line width=0.5pt},
        darrow/.style={-Latex, line width=0.5pt, dashed}
    ]
        \node[font=\footnotesize\itshape] at (2.3,2.95) {A. Molecular representations};
        \node[font=\footnotesize\itshape] at (7.1,2.95) {B. Kernel construction};
        \node[font=\footnotesize\itshape] at (12.0,2.95) {C. Nested evaluation and audit};

        \node[block, minimum width=1.30cm] (smiles) at (0,0.10) {SMILES\\strings};
        \node[block, minimum width=2.45cm] (morgan) at (2.75,1.45) {Morgan fingerprint\\Tanimoto kernel $K_M$};
        \node[block, minimum width=2.45cm] (desc) at (2.75,-1.20) {RDKit, Mordred, Deep-PK\\fold-local components $z_i$};

        \node[block, minimum width=2.35cm] (qkernel) at (5.95,-1.20) {Quantum map\\fidelity kernel $K_Q$};
        \node[block, minimum width=2.35cm] (ckernel) at (5.95,-2.60) {Classical controls\\$K_C^{\rm lin}, K_C^{\rm rbf}$};
        \node[block, minimum width=3.15cm] (mix) at (8.35,1.45) {MKL combination\\$K_\alpha=(1-\alpha)K_M+\alpha K_D$\\$K_D\in\{K_Q,K_C\}$};
        \node[block, minimum width=1.70cm] (svc) at (11.20,1.45) {SVM classifier\\inner folds\\select $C,\alpha$};
        \node[block, minimum width=2.55cm] (pred) at (13.75,1.45) {Held-out molecules\\metrics, molecule audit,\\cliff and seed analyses};

        \draw[arrow] (smiles.east) -- (morgan.west);
        \draw[arrow] (smiles.east) -- (desc.west);
        \draw[arrow] (morgan.east) -- (mix.west);
        \draw[arrow] (desc.east) -- (qkernel.west);
        \draw[arrow] (qkernel.north) |- (mix.south);
        \draw[darrow] (desc.south east) |- (ckernel.west);
        \draw[darrow] (ckernel.east) -| (mix.south);
        \draw[arrow] (mix.east) -- (svc.west);
        \draw[arrow] (svc.east) -- (pred.west);
    \end{tikzpicture}}
    \caption{QBioFusion-QSAR workflow. Simplified molecular-input line-entry system (SMILES) strings are mapped through a Morgan/Tanimoto branch and a descriptor branch. The descriptor branch is fitted inside each training partition before constructing the quantum fidelity kernel $K_Q$ and matched classical descriptor controls $K_C$. Multiple-kernel weights are selected within outer-training folds and interpreted through held-out, activity-cliff, curation, repeated-seed, and molecule-level audits.}
    \label{fig:pipeline}
\end{figure*}

\section{Background and Problem Setting}

The target task is binary classification. Each molecule $i$ is represented by a simplified molecular-input line-entry system (SMILES) string and a label $y_i \in \{0,1\}$ indicating the curated hallucinogenic activity class. The positive class is small, so evaluation uses balanced accuracy, Matthews correlation coefficient (MCC), $F_1$ score, sensitivity, specificity, area under the receiver-operating-characteristic curve (ROC-AUC), area under the precision-recall curve (PR-AUC), and Brier score. MCC is emphasized because it summarizes all four entries of the confusion matrix and is informative under class imbalance.

The classical baseline uses Morgan fingerprints, following the extended-connectivity fingerprint family widely used in quantitative structure-activity relationship modeling and virtual screening \cite{Rogers2010}. Let $m_i \in \{0,1\}^{1024}$ denote the radius-3 Morgan fingerprint of molecule $i$. The Tanimoto kernel is
\begin{equation}
    \KM(i,j)=\frac{m_i^\top m_j}{\lVert m_i\rVert_1+\lVert m_j\rVert_1-m_i^\top m_j},
    \label{eq:tanimoto}
\end{equation}
with the convention that the denominator is nonzero for the molecules considered. This kernel encodes the common medicinal-chemistry assumption that similar substructure fingerprints often imply similar biological behavior. Activity cliffs challenge that assumption: two molecules can be highly similar by a fingerprint and nevertheless differ in activity. Dedicated activity-cliff evaluation is therefore useful in molecular machine learning \cite{VanTilborg2022}.

The descriptor block concatenates numerical RDKit, Mordred, and Deep-PK columns supplied with the source data \cite{LandrumRDKit,Moriwaki2018,Myung2024}. Deep-PK features are treated as externally generated pharmacokinetic and toxicity covariates; they are kept fixed, and PsychLight-A labels are not used to train or tune them in this study. In each training partition, descriptor values are imputed, scaled, reduced to $q$ principal components with $q=4$, and rescaled to $[-\pi,\pi]$. The resulting four-dimensional vector for molecule $i$ is denoted $z_i\in\mathbb{R}^{4}$. When the descriptor source must be explicit, $z_i^{R}$, $z_i^{M}$, $z_i^{D}$, and $z_i^{U}$ denote fold-local components from RDKit, Mordred, Deep-PK, and their union, respectively. Four components were used to match the four-qubit feature map; in the seed-42 outer folds, the union components captured $0.548\pm0.013$ of the post-filtering training-fold descriptor variance. The implementation uses scikit-learn for preprocessing and support vector classification \cite{Pedregosa2011}. Descriptor preprocessing is fitted on the relevant training partition before being applied to held-out molecules.

\section{Method}

Fig.~\ref{fig:pipeline} summarizes the method. The first kernel is the Morgan/Tanimoto kernel in (\ref{eq:tanimoto}). The second is a quantum fidelity kernel over the reduced descriptor representation. Unless a descriptor source is shown explicitly, $z_i$ denotes the active fold-local vector, and the main QMKL model uses the union vector $z_i^{U}$. The data-encoding circuit follows an instantaneous quantum polynomial-time (IQP)-style feature map. Given $z_i \in \R^q$, the circuit prepares
\begin{align}
    |\phi(z_i)\rangle &= 2^{-q/2}\sum_{b\in\{0,1\}^{q}}
    \exp(i\theta_i(b))|b\rangle, \\
    \theta_i(b) &= \sum_{r=1}^{q}b_r z_{ir}
    +\lambda\sum_{r<s} b_r b_s z_{ir}z_{is},
    \label{eq:state}
\end{align}
with entangling scale $\lambda=0.5$. For descriptor source $s$, the quantum descriptor kernel is
\begin{equation}
    K_Q^{s}(i,j)=\left|\langle \phi(z_i^{s})|\phi(z_j^{s})\rangle\right|^2.
    \label{eq:qkernel}
\end{equation}
The main model uses $K_Q^{U}$ and writes it as $\KQ$ for compactness; a source-specific RDKit control uses $K_Q^{R}$.
Equivalently, if $\rho_i=|\phi(z_i)\rangle\langle\phi(z_i)|$, then $\KQ(i,j)=\mathrm{Tr}(\rho_i\rho_j)$. The kernel was computed by exact statevector simulation of this four-qubit feature map; no quantum hardware or finite-shot sampling was used. The quantum content studied here is therefore the fidelity geometry induced by the feature map, not a hardware-execution advantage. The scale $\lambda=0.5$ was fixed for all experiments rather than tuned on PsychLight-A. The descriptor branch is therefore a fidelity geometry over molecules: two molecules are close when their descriptor-derived quantum states have high overlap.

The multiple-kernel model uses the convex combination
\begin{equation}
    \Kmix(i,j)=(1-\alpha_Q)\KM(i,j)+\alpha_Q\KQ(i,j),
    \qquad 0\leq \alpha_Q\leq 1.
    \label{eq:mixed}
\end{equation}
The convex combination has a standard multiple-kernel interpretation \cite{Scholkopf2002,Gonen2011}. If $\Psi_M$ and $\Psi_Q$ are feature maps associated with $\KM$ and $\KQ$, then
\begin{equation}
    \Psi_\alpha(x)=\left[\sqrt{1-\alpha_Q}\,\Psi_M(x),
    \sqrt{\alpha_Q}\,\Psi_Q(x)\right]
    \label{eq:directsum}
\end{equation}
induces $\Kmix(i,j)=\langle\Psi_\alpha(x_i),\Psi_\alpha(x_j)\rangle$. The SVM therefore optimizes one margin in a direct-sum feature space whose first component is substructure similarity and whose second component is descriptor-state fidelity. With labels encoded as $\tilde y_i\in\{-1,+1\}$ and class weights $w_{y_i}$, the corresponding soft-margin objective is
\begin{equation}
\begin{aligned}
    \min_{f,b,\xi}\, &\frac{1}{2}\lVert f\rVert_{\mathcal H_\alpha}^2
    +C\sum_{i\in\mathcal T} w_{y_i}\xi_i,\\
    \mathrm{s.t.}\, &(\tilde y_i)(f(i)+b)\geq 1-\xi_i,\qquad \xi_i\geq0 .
\end{aligned}
\label{eq:svm_obj}
\end{equation}
This formulation is appropriate for the small-data setting because the quantum circuit supplies a fixed kernel and the selected degrees of freedom remain limited to the SVM regularization and the kernel-combination weight.

The selected classifier is a support vector machine with a precomputed kernel and balanced class weights. In the learned-alpha variant, candidate values are searched on a simplex grid. For the two-kernel case, the candidate set is $\alpha_Q\in\{0,0.25,0.50,0.75,1.00\}$. Fixed-alpha profiling evaluates $\alpha_Q \in \{0,0.05,0.10,0.25,0.50,0.75,1.00\}$ at $C=10$. This separation between learned-alpha selection and fixed-alpha profiling makes the analysis interpretable: the learned model measures the selected operating point, whereas the fixed profile shows how performance changes as descriptor-kernel weight increases from a Morgan-only classifier to a classifier with high descriptor-kernel weight.

All model selection is performed inside the training data available to the fold. For an outer split $r$, let $\mathcal{T}_r$ be the outer-training molecules and $\mathcal{S}_r$ the held-out molecules. Candidate values of $C$ and $\alpha$ are selected only by inner folds inside $\mathcal{T}_r$; the selected pipeline is then refit on $\mathcal{T}_r$ and applied once to $\mathcal{S}_r$. Imputation, scaling, dimensionality reduction, feature rescaling, kernel computation, and SVM fitting are recomputed within the relevant training portion before scoring held-out molecules. This protocol gives every held-out molecule a prediction from a model whose preprocessing and hyperparameters were estimated from the corresponding training partition only. The practical comparison uses Morgan/Tanimoto with $C=1$ and QMKL with $C=10$. A matched-regularization comparison also evaluates Morgan/Tanimoto with $C=10$ against QMKL with $C=10$.

Classical descriptor-kernel controls use the same four fold-local descriptor principal components as the quantum kernel. In addition to the full RDKit--Mordred--Deep-PK descriptor union, source-specific controls fit $z_i^{R}$, $z_i^{M}$, or $z_i^{D}$ from RDKit-only, Mordred-only, or Deep-PK-only columns. The RDKit-specific quantum control $K_Q^{R}$ is also reported because it tests whether the observed QMKL behavior is already present when the quantum fidelity map receives only RDKit descriptor components. These controls are essential because an improvement from $\KQ$ can arise either from the descriptor information entering the model or from the quantum fidelity geometry used to compare those descriptors. The descriptor-only controls use either the shifted normalized linear kernel
\begin{equation}
    K_C^{\mathrm{lin}}(i,j)=\frac{1}{2}\left(1+
    \frac{z_i^\top z_j}{\lVert z_i\rVert_2\lVert z_j\rVert_2}\right)
    \label{eq:linearcontrol}
\end{equation}
or the median-heuristic radial basis-function kernel
\begin{equation}
    K_C^{\mathrm{rbf}}(i,j)=\exp\{-\gamma\lVert z_i-z_j\rVert_2^2\},
    \label{eq:rbfcontrol}
\end{equation}
where $\gamma$ is chosen from a multiplier grid around the training-fold median heuristic. The Morgan-plus-descriptor controls use $K=(1-\alpha)\KM+\alpha\KC$ with the same $C$ grid, descriptor-weight grid, inner-fold selection, and outer folds as the QMKL model. This matched-control design isolates the role of quantum fidelity geometry from the role of adding reduced continuous descriptor information.

A molecule-level discordance audit characterizes how the descriptor kernel changes held-out classifications. Let $\hat y_M(i)$ and $\hat y_Q(i)$ be the out-of-fold predictions from the Morgan and QMKL models. Each molecule is assigned to one of four categories: Morgan-wrong/QMKL-correct, Morgan-correct/QMKL-wrong, both correct, or both wrong. The score shift is summarized as $\Delta p_i=p_Q(y_i|x_i)-p_M(y_i|x_i)$, where $p_Q$ and $p_M$ are logistic transformations of the out-of-fold SVM decision values assigned to the true class. These values are used only to describe how close a molecule is to the decision boundary; the discrete held-out predictions and decision-boundary audit are treated as the primary evidence.

Activity-cliff regions are defined through opposite-class Morgan similarity. For molecule $i$,
\begin{equation}
    A_i = \max_{j: y_j\neq y_i}\KM(i,j).
\end{equation}
A molecule is assigned to the activity-cliff region when $A_i\geq \tau$. The main threshold is $\tau=0.6$; sensitivity checks use $\tau=0.5$ and $\tau=0.7$. Curation sensitivity is evaluated with chiral Morgan fingerprints and connectivity-group-aware folds, where molecules sharing the same International Chemical Identifier key (InChIKey) connectivity block are kept in the same outer fold.

\section{Experimental Design}

The study uses PsychLight-A as the primary benchmark. Table~\ref{tab:dataset} summarizes the experimental inputs. The uploaded processed descriptor union contains 1913 numerical columns after removing target and metadata fields. The descriptor matrix is never reduced once globally. Instead, each outer or inner training partition learns its own imputation, scaling, principal-component basis, and $[-\pi,\pi]$ rescaling, so the quantum coordinates $z_i$ are fold-dependent training estimates. The original audit flags one exact canonical-SMILES duplicate pair, 5-HT and Bufotenin, and same-connectivity stereochemical groups including R-MDA and S-MDA with opposite labels. These rows are retained for the primary benchmark and evaluated again in the curation-sensitivity analysis.

\begin{table}[t]
\centering
\caption{Dataset and molecular representation summary. Outer folds are held-out evaluation folds; inner folds are used only for model and kernel-weight selection inside the corresponding outer-training set.}
\label{tab:dataset}
\begin{tabular}{ll}
\toprule
Property & Value \\
\midrule
Dataset & PsychLight-A \\
Molecules & 54 \\
Positive / negative labels & 18 / 36 \\
Classical input & Morgan radius 3, 1024 bits \\
Classical kernel & Tanimoto similarity \\
Descriptor input & RDKit + Mordred + Deep-PK \\
Descriptor dimension & 4 fold-local principal components \\
Quantum kernel & Descriptor fidelity kernel \\
Outer / inner folds & 5 / 2 \\
Main cliff threshold & $A_i\geq 0.6$ \\
\bottomrule
\end{tabular}
\end{table}

The primary classifiers are Morgan/Tanimoto SVM and quantum descriptor QMKL SVM. The primary evaluation is not a single train/test split: it is one stratified five-fold out-of-fold protocol with random state 42, so each molecule is scored once as held-out data. The practical performance comparison is Morgan/Tanimoto with $C=1$ against QMKL with $C=10$. The matched comparison is Morgan/Tanimoto with $C=10$ against QMKL with $C=10$. Fixed-alpha profiling is performed with $C=10$. Repeated random-state analysis repeats the same five-fold protocol over seeds 40--49 to quantify partition sensitivity and selected kernel weights. Classical descriptor-kernel controls are evaluated on the same seed-42 outer folds and use inner-fold selection over $C\in\{1,10\}$, descriptor weight $\alpha\in\{0,0.05,0.10,0.25,0.50,0.75,1.00\}$, and RBF median-heuristic multipliers $\{0.25,1,4\}$.

The validation design is deliberately centered on comparability rather than model complexity. Every model sees the same outer-test molecules in each fold. Every descriptor-derived kernel, quantum or classical, receives the same fold-local four-dimensional descriptor representation. The matched-$C$ comparison fixes the SVM regularization to expose the effect of replacing or adding the descriptor kernel, whereas the practical comparison reports the prespecified primary operating points. The repeated-seed study provides the partition-dependence context needed to characterize a small-data QML method. An anonymous reproducibility package accompanies the submission and contains the scripts, figures, and tables used for the reported analyses.

The practical and matched comparisons serve different interpretive purposes. The practical comparison reports the primary prespecified operating points used for the main seed-42 evaluation. The matched-$C$ comparison fixes the regularization parameter so that changes in held-out predictions can be attributed to the MKL combination rather than to a different margin penalty. The fixed-alpha profile is diagnostic rather than a second model-selection procedure: it evaluates the same outer folds while sweeping descriptor-kernel weight to show whether performance is improved by small residual weights or by large descriptor-kernel weights.

\section{Results}

Table~\ref{tab:primary} reports representation-only controls followed by Morgan-anchored MKL comparisons. It also defines the table metrics: ACC is accuracy, BACC is balanced accuracy, MCC is Matthews correlation coefficient, $F_1$ balances precision and recall, and TP/FP/FN/TN are confusion-matrix counts. Morgan/Tanimoto reached 0.815 accuracy and 0.613 MCC, whereas RDKit-only, Mordred-only, Deep-PK-only, and union fold-local component classifiers reached MCC values of 0.373, 0.368, 0.219, and 0.185. Thus the reduced descriptor components $z_i$ do not replace fingerprint similarity by themselves.

Adding the union quantum descriptor kernel changed the primary result: QMKL increased ACC from 0.815 to 0.833, BACC from 0.819 to 0.833, MCC from 0.613 to 0.645, and $F_1$ from 0.750 to 0.769. In the matched-$C$ comparison, QMKL increased MCC from 0.548 to 0.645 and reduced false negatives from five to three while maintaining six false positives. The RDKit-specific quantum control $K_Q^{R}$ reached MCC 0.515, below Morgan/Tanimoto and union QMKL, so the favorable primary result is not explained by RDKit-only quantum components.

Classical descriptor-kernel controls constrain the interpretation. Nested Morgan-plus-linear and Morgan-plus-RBF controls reached MCC values of 0.503 and 0.564, below learned QMKL and the Morgan $C=1$ point. Fixed-alpha profiling showed that, at $C=10$ and $\alpha=0.25$, the classical linear descriptor combination reached the same threshold metrics as fixed small-weight QMKL (ACC 0.815, BACC 0.806, MCC 0.597). Thus part of the threshold-dependent change is shared with small continuous descriptor contributions, while learned union QMKL gives the strongest seed-42 result among the tested descriptor-kernel combinations.

\begin{table}[t]
\centering
\caption{Seed-42 five-fold evaluation. ACC is accuracy, BACC balanced accuracy, MCC Matthews correlation coefficient, and TP/FP/FN/TN confusion-matrix counts. Rows labeled $z_i$ use descriptor-only SVMs; $K_Q^U$ and $K_Q^R$ denote union and RDKit-only quantum fidelity kernels.}
\label{tab:primary}
\scriptsize
\setlength{\tabcolsep}{1.6pt}
\begin{tabular}{lcccccc}
\toprule
Model & ACC & BACC & MCC & $F_1$ & TP/FP & FN/TN \\
\midrule
\multicolumn{7}{l}{\emph{Single representations}} \\
Morgan $C=1$ & .815 & .819 & .613 & .750 & 15/7 & 3/29 \\
RDKit $z_i$ & .704 & .694 & .373 & .600 & 12/10 & 6/26 \\
Mordred $z_i$ & .685 & .694 & .368 & .605 & 13/12 & 5/24 \\
Deep-PK $z_i$ & .648 & .611 & .219 & .486 & 9/10 & 9/26 \\
Union $z_i$ & .611 & .597 & .185 & .488 & 10/13 & 8/23 \\
\midrule
\multicolumn{7}{l}{\emph{Morgan-anchored combinations}} \\
QMKL $K_Q^U$ $C=10$ & .833 & .833 & .645 & .769 & 15/6 & 3/30 \\
QMKL $K_Q^R$ $C=10$ & .778 & .764 & .515 & .684 & 13/7 & 5/29 \\
Morgan $C=10$ & .796 & .778 & .548 & .703 & 13/6 & 5/30 \\
QMKL $\alpha_Q\geq.25$ & .815 & .806 & .597 & .737 & 14/6 & 4/30 \\
Morgan+linear & .759 & .764 & .503 & .683 & 14/9 & 4/27 \\
Morgan+RBF & .796 & .792 & .564 & .718 & 14/7 & 4/29 \\
\bottomrule
\end{tabular}
\end{table}

The molecule-level audit in Table~\ref{tab:molaudit} localizes the matched-$C$ change. N-Me-5-HT and N-Me-tryptamine, both positive tryptamine-like molecules near the Morgan boundary, were false negative with Morgan/Tanimoto and true positive with QMKL. Their logistic-transformed decision scores moved from 0.463 to 0.517 and from 0.497 to 0.535, respectively. No reverse discordant case was observed. The two-sided exact McNemar test was $p=0.50$, and stratified paired bootstrap intervals included zero for ACC, BACC, MCC, and $F_1$; the direction is favorable, but the 54-molecule sample limits statistical resolution.

\begin{table}[t]
\centering
\caption{Molecule-level matched-$C$ audit. FN$\rightarrow$TP denotes a Morgan false-negative (FN) classified as true positive (TP) by QMKL. $p_M$ and $p_Q$ are logistic-transformed out-of-fold SVM decision values for the positive class.}
\label{tab:molaudit}
\begin{tabular}{lcccc}
\toprule
Molecule/category & $y$ & $p_M$ & $p_Q$ & Change/count \\
\midrule
N-Me-5-HT & 1 & 0.463 & 0.517 & FN$\rightarrow$TP \\
N-Me-tryptamine & 1 & 0.497 & 0.535 & FN$\rightarrow$TP \\
\midrule

\end{tabular}
\end{table}

Table~\ref{tab:cliff} reports activity-cliff sensitivity. At $\tau=0.5$, QMKL increased MCC from 0.037 to 0.241 over 21 molecules; at the main threshold $\tau=0.6$, it increased MCC from 0.071 to 0.220 over 13 molecules. The $\tau=0.7$ subset contained only two molecules and is too small for a stable conclusion. The analysis supports the interpretation that the descriptor kernel changes decisions in locally ambiguous Morgan neighborhoods.

\begin{table}[t]
\centering
\caption{Matched-$C$ activity-cliff threshold sensitivity. The threshold $\tau$ is applied to $A_i$, the maximum opposite-class Morgan/Tanimoto similarity; $n$ is the number of molecules in the cliff subset.}
\label{tab:cliff}
\begin{tabular}{lccccc}
\toprule
$\tau$ / model & $n$ & ACC & BACC & MCC & $F_1$ \\
\midrule
0.5 Morgan & 21 & 0.524 & 0.518 & 0.037 & 0.583 \\
0.5 QMKL & 21 & 0.619 & 0.609 & 0.241 & 0.692 \\
0.6 Morgan & 13 & 0.538 & 0.536 & 0.071 & 0.571 \\
0.6 QMKL & 13 & 0.615 & 0.607 & 0.220 & 0.667 \\
\bottomrule
\end{tabular}
\end{table}

\FloatBarrier

\begin{figure}[t]
    \centering
    \begin{tikzpicture}
    \begin{axis}[
        width=\linewidth,
        height=0.62\linewidth,
        xlabel={Descriptor-kernel weight $\alpha$},
        ylabel={MCC},
        xmin=-0.03, xmax=1.03,
        ymin=0.15, ymax=0.66,
        grid=both,
        tick label style={font=\scriptsize},
        label style={font=\footnotesize},
        legend style={font=\scriptsize, at={(0.03,0.03)}, anchor=south west, draw=none, fill=none},
        legend cell align={left}
    ]
    \addplot[oiBlue, very thick, solid, mark=*, mark options={solid, fill=oiBlue}] coordinates {(0,0.548) (0.05,0.597) (0.10,0.597) (0.25,0.597) (0.50,0.515) (0.75,0.403) (1.00,0.424)};
    \addlegendentry{Quantum descriptor}
    \addplot[oiVermillion, very thick, dashed, mark=square*, mark options={solid, fill=oiVermillion}] coordinates {(0,0.548) (0.05,0.548) (0.10,0.548) (0.25,0.597) (0.50,0.548) (0.75,0.515) (1.00,0.190)};
    \addlegendentry{Linear descriptor}
    \addplot[oiGreen, very thick, densely dotted, mark=triangle*, mark options={solid, fill=oiGreen}] coordinates {(0,0.548) (0.05,0.548) (0.10,0.548) (0.25,0.500) (0.50,0.548) (0.75,0.483) (1.00,0.341)};
    \addlegendentry{RBF descriptor}
    \end{axis}
    \end{tikzpicture}
    \caption{Fixed-alpha Matthews correlation coefficient (MCC) profile at $C=10$ for the quantum fidelity kernel and matched classical descriptor kernels. The pure Morgan point is $\alpha=0$.}
    \label{fig:alpha}
\end{figure}
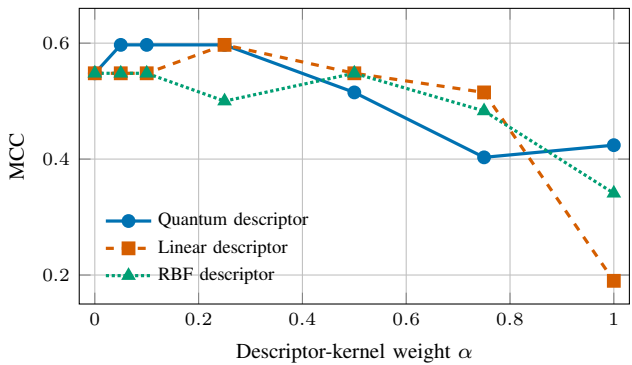

Fig.~\ref{fig:alpha} compares fixed-alpha profiles. Small quantum weights improved threshold-dependent metrics relative to the pure Morgan $C=10$ setting, whereas large quantum weights reduced performance. The linear descriptor kernel showed a similar small-weight effect at $\alpha=0.25$, and the RBF kernel was strongest at the Morgan-only point.

Table~\ref{tab:curation} summarizes curation sensitivity. Achiral and chiral Morgan fingerprints gave identical ordinary seed-42 metrics at $C=1$. Connectivity-group-aware folds reduced MCC from 0.613 to 0.583 at $C=1$, whereas group-aware chiral Morgan at $C=10$ recovered the ordinary $C=1$ metrics. Thus duplicate and stereochemical edge cases matter, while the matched-$C$ QMKL changes remain localized to the two boundary-near tryptamine-like molecules.

\begin{table}[t]
\centering
\caption{Morgan chirality and connectivity-group sensitivity. Ordinary splits are standard stratified folds; group-aware splits keep molecules with the same InChIKey connectivity block in the same outer fold.}
\label{tab:curation}
\begin{tabular}{lcccc}
\toprule
Setting & $C$ & ACC & BACC & MCC \\
\midrule
Ordinary achiral & 1 & 0.815 & 0.819 & 0.613 \\
Ordinary chiral & 1 & 0.815 & 0.819 & 0.613 \\
Group-aware achiral & 1 & 0.796 & 0.806 & 0.583 \\
Group-aware chiral & 1 & 0.796 & 0.806 & 0.583 \\
Group-aware chiral & 10 & 0.815 & 0.819 & 0.613 \\
\bottomrule
\end{tabular}
\end{table}

Repeated random-state analysis over seeds 40--49 quantified partition dependence. The mean selected quantum weight was 0.27, and $\alpha_Q$ was nonzero in 22 of 50 outer folds. Mean MCC was $0.540\pm0.077$ for Morgan/Tanimoto and $0.487\pm0.124$ for learned QMKL. The primary five-fold result therefore shows a promising improvement under one complete held-out partitioning, while the ten-partition analysis does not establish stable average superiority at this sample size.

\section{Discussion and Conclusion}

The experiments answer the two research questions at different levels of evidence. The primary five-fold evaluation indicates that an embedded quantum descriptor kernel can improve threshold-dependent classification on PsychLight-A: accuracy increased from 0.815 to 0.833 and MCC from 0.613 to 0.645. The matched-$C$ audit localizes the change to N-Me-5-HT and N-Me-tryptamine, two positive tryptamine-like ligands near the Morgan decision boundary whose predictions changed from false negative to true positive. The activity-cliff analysis supports the same interpretation.

The representation and kernel controls define the scope. RDKit, Mordred, Deep-PK, and union descriptor components do not match Morgan/Tanimoto when used alone. Morgan-plus-classical descriptor kernels and the RDKit-only quantum control $K_Q^{R}$ also remain below learned union QMKL in the seed-42 comparison, although the fixed small-weight linear control reproduces the fixed small-weight QMKL threshold change. The result is therefore a controlled residual-kernel finding: small descriptor information is useful, and learned quantum-fidelity MKL gives the strongest seed-42 threshold-dependent result among the tested descriptor-kernel combinations.

The repeated random-state result gives the appropriate scope. Learned QMKL is partition-sensitive, and its mean MCC over seeds 40--49 is below the Morgan/Tanimoto mean. 

The repeated random-state result defines the appropriate scope of the finding. Learned QMKL is partition-sensitive, and its mean MCC over seeds 40--49 is below the Morgan/Tanimoto mean. Together with the molecule-level audit, this reframes the evidence as localized rather than uniformly average-improving: the embedded quantum descriptor kernel corrected selected boundary-near molecules in the primary held-out partitioning, but did not establish stable superiority across random folds at this sample size.

QBioFusion-QSAR therefore contributes an auditable QMKL-QSAR protocol for small-data ligand classification. The framework evaluates aggregate held-out performance, matched classical descriptor-kernel controls, activity-cliff sensitivity, curation sensitivity, repeated partitioning, and the individual molecules responsible for decision-boundary changes. On PsychLight-A, the strongest evidence is a localized residual-similarity effect in which the Morgan-anchored quantum fidelity kernel corrected two positive tryptamine-like ligands and improved activity-cliff subset MCC. Future work should test this protocol on larger, independently curated, stereochemistry-aware, and activity-cliff-enriched ligand datasets. The main contribution for QAI is not a broad quantum-advantage claim, but a reproducible framework for determining when and where quantum kernels add useful residual similarity in small-data QSAR.

\begingroup
\fontsize{6.4}{6.55}\selectfont

\endgroup

\end{document}